\documentclass[%
reprint,
superscriptaddress,
amsmath,amssymb,amsthm,mathrsfs,
aps,
longbibliography,
prl,
]{revtex4-1}
\usepackage{graphicx}
\usepackage{dcolumn}
\usepackage{bm}
\usepackage[usenames]{color}
\usepackage[ 
margin=0.75in,
]{geometry}

\def\NMO{NaMnO$_2$}
\def\CMO{CuMnO$_2$}

\def\TN{$T_{\mathrm{N}}$}

\begin{document}

\title{
Nanoscale Degeneracy Lifting in a Geometrically Frustrated Antiferromagnet
}

\author{Benjamin A. Frandsen}
\email{benfrandsen@byu.edu}
	\affiliation{ %
	Department of Physics and Astronomy, Brigham Young University, Provo, Utah 84602, USA.
} %

\author{Emil S. Bozin}
\email{bozin@bnl.gov}
\affiliation{%
	Condensed Matter Physics and Materials Science Division, Brookhaven National Laboratory, Upton, NY 11973, USA.
}%

\author{Eleni Aza}
\affiliation{
	Institute of Electronic Structure and Laser, Foundation for Research and Technology---Hellas, Vassilika Vouton, 71110 Heraklion, Greece.
}
\affiliation{
	Department of Materials Science and Engineering, University of Ioannina, 451 10 Ioannina, Greece.
}

\author{Antonio Fern\'andez Mart\'inez}
\affiliation{
	Institute of Electronic Structure and Laser, Foundation for Research and Technology---Hellas, Vassilika Vouton, 71110 Heraklion, Greece.
}

\author{Mikhail Feygenson}
\affiliation{
	Neutron Scattering Division, Oak Ridge National Laboratory, Oak Ridge, Tennessee 37831, USA.
}%
\affiliation{
	J\"ulich Centre of Neutron Science, Forschungszentrum J\"ulich, 52428, J\"ulich Germany.
}%

\author{Katharine Page}
\affiliation{
	Neutron Scattering Division, Oak Ridge National Laboratory, Oak Ridge, Tennessee 37831, USA.
}%
\affiliation{
	Materials Science and Engineering Department, University of Tennessee, Knoxville, Tennessee 37996, USA.
}%

\author{Alexandros Lappas}
\email{lappas@iesl.forth.gr}
\affiliation{
Institute of Electronic Structure and Laser, Foundation for Research and Technology---Hellas, Vassilika Vouton, 71110 Heraklion, Greece.
}

\begin{abstract}
The local atomic and magnetic structures of the compounds $A$MnO$_2$ ($A$ = Na, Cu), which realize a geometrically frustrated, spatially anisotropic triangular lattice of Mn spins, have been investigated by atomic and magnetic pair distribution function analysis of neutron total scattering data. Relief of frustration in CuMnO$_2$ is accompanied by a conventional cooperative symmetry-lowering lattice distortion driven by N\'eel order. In NaMnO$_2$, however, the distortion has a short-range nature. A cooperative interaction between the locally broken symmetry and short-range magnetic correlations lifts the magnetic degeneracy on a nanometer length scale, enabling long-range magnetic order in the Na-derivative. The degree of frustration, mediated by residual disorder, contributes to the rather differing pathways to a single, stable magnetic ground state in these two related compounds. This study demonstrates how nanoscale structural distortions that cause local-scale perturbations can lift the ground state degeneracy and trigger macroscopic magnetic order. \footnote{Notice: This manuscript has been co-authored by UT-Battelle, LLC, under contract DE-AC05-00OR22725 with the US Department of Energy (DOE). The US government retains and the publisher, by accepting the article for publication, acknowledges that the US government retains a nonexclusive, paid-up, irrevocable, worldwide license to publish or reproduce the published form of this manuscript, or allow others to do so, for US government purposes. DOE will provide public access to these results of federally sponsored research in accordance with the DOE Public Access Plan (http://energy.gov/downloads/doe-public-access-plan).}
\end{abstract}

\maketitle

\section{Introduction}

In geometrically frustrated materials, the spatial arrangement of magnetic moments prevents the simultaneous satisfaction of competing magnetic interactions~\cite{greed;jmc01}. This hinders the formation of magnetic order and may occasionally lead to exotic, macroscopically degenerate spin-liquid ground states for which magnetic order is absent even at zero temperature~\cite{balen;n10,lee;jpsj10}. In the more common case where order develops at finite temperature, the transition mechanism itself can be quite unusual, often originating from extreme sensitivity to small perturbations resulting from thermal fluctuations, single-ion anisotropy, Dzyaloshinskii-Moriya and dipolar interactions, disorder, magnetoelastic coupling, and more~\cite{moess;pt06}. Coupling between magnetic and other degrees of freedom can lift the intrinsic degeneracies and even support novel phenomena beyond magnetism, such as magnetoelectric states established by uniform lattice distortions~\cite{lee;n08} and domain-wall-driven competing metastable microphases~\cite{kamiy;prl12}. 

An added layer of complexity may arise when frustrated geometries occur in strongly correlated transition metal oxide (TMO) systems. The possibility of stabilizing unusual, spatially inhomogeneous ground states through competition among simultaneously active electronic degrees of freedom (i.e. spin, charge, and orbital) has been well documented in TMOs~\cite{dagot;s05,sheno;cpc06}. Combining this affinity for spatial inhomogeneity with geometrical frustration may produce novel behavior. 

\begin{figure*}
	\includegraphics[width=140mm]{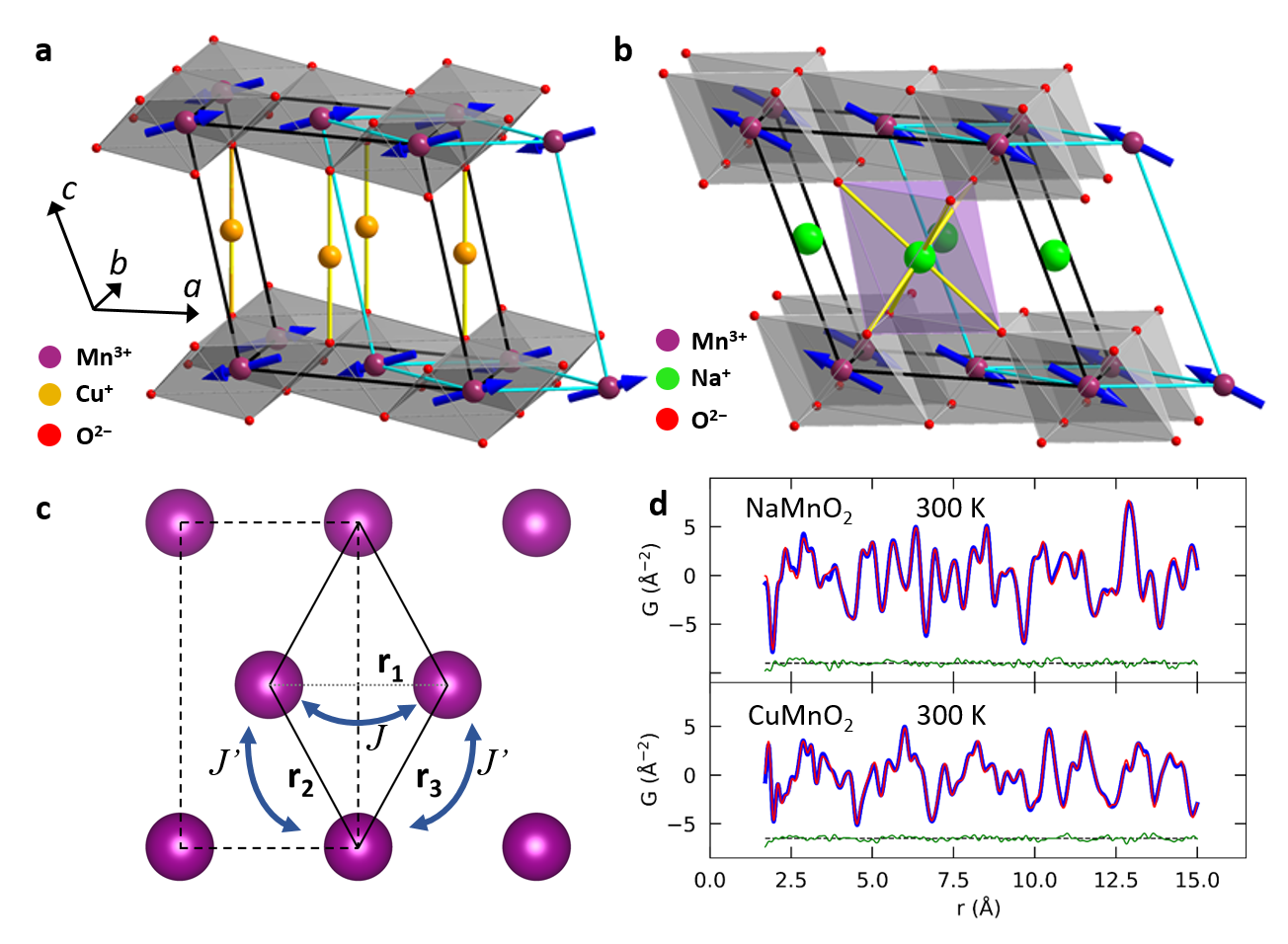}
	\caption{\label{fig:structure} \textbf{Structure of \CMO\ and $\alpha$-\NMO.} (a) Atomic and magnetic structures of \CMO\ and (b) $\alpha$-\NMO. The distortion of the monoclinic unit cell (black) to triclinic cell (blue) is depicted. The arrows depict the three-dimensional antiferromagnetic spin configurations, corresponding to propagation vectors (in the triclinic setting) of $\mathbf{k}=(0,1/2,/1/2)$ for CMO~\cite{vecch;prb10} and $\mathbf{k}=(0,1/2,0)$ for NMO~\cite{giot;prl07}. (c) Projection onto the basal plane of the triangular Mn$^{3+}$ sublattice. Dashed and solid lines show the monoclinic and triclinic unit cells, respectively. (d) PDF patterns of \NMO\ (top) and \CMO\ (bottom) collected at room temperature. The blue curves represent the data, the red curves the atomic PDF fits using the monoclinic structural model, and the lower green curves the fit residuals, offset for clarity.}
	
\end{figure*}

The insulating sister compounds $\alpha$-\NMO\ (NMO)~\cite{paran;jssc71} and \CMO\ (CMO)~\cite{topfe;zk95} provide a valuable testbed for exploring geometrically frustrated magnetism in strongly correlated TMOs. The $A$MnO$_2$ crystal structures [Fig.~\ref{fig:structure}(a, b)] entail spacer layers of primarily monovalent $A$ = Na or Cu cations that separate layers of edge-sharing MnO$_6$ octahedra.
The Jahn-Teller active, antiferromagnetically (AF) coupled Mn$^{3+}$ ions ($S=2$) form an anisotropic triangular network, embodying a simple example of a frustrated topology imposed on the two-dimensional Heisenberg model.  At ambient conditions, the triangles are isosceles with $r_1 < r_2 = r_3$, as shown in Fig.~\ref{fig:structure}(c). Consequently, the exchange interactions $J$ (along $r_1$) and $J'$ (along $r_2$ and $r_3$) become nondegenerate, with $J'/J < 1$. The frustration of the triangular lattice is therefore partially relieved. Nevertheless, the magnetic interactions along the two isosceles legs remain degenerate, and the resulting partial frustration of the AF interactions remains important even though these systems would not be considered ``highly frustrated'' magnets. Indeed, the magnetic frustration has been implicated in numerous unusual properties displayed by these systems, such as quasi-one-dimensional magnetic interactions~\cite{stock;prl09} and rare Higgs-type excitations~\cite{dally;nc18} in NMO. In general, systems with imperfect frustration, e.g. the anisotropic triangular lattice considered here, may still exhibit rich behavior such as ordering by disorder and Peierls-type transitions~\cite{becca;prl02}.

In CMO, magnetic long-range order (LRO)~\cite{damay;prb09} develops below \TN\ = 65~K. This is accomplished through a concomitant monoclinic-to-triclinic structural phase transition that further distorts the triangles such that $r_2 \ne r_3$, fully lifting the degeneracy of the triangular lattice and selecting a unique magnetic ground state~\cite{vecch;prb10}. The ordered state exhibits AF alignment along the shorter formerly isosceles leg and ferromagnetic alignment along the other formerly isosceles leg [Fig.~\ref{fig:structure}(a)], with antiferromagnetic coupling between layers~\cite{damay;prb09}. The energy gain from relieving the magnetic frustration outweighs the elastic energy cost of the triclinic distortion~\cite{zorko;sr15}; hence, this effect is magnetically driven. Fig.~\ref{fig:structure}(c) displays the projection of the monoclinic and triclinic unit cells on the Mn layer.

Although NMO is more frustrated than CMO ($J'/J = 0.44$~\cite{zorko;prb08} compared to 0.27 for CMO~\cite{zorko;sr15}, resulting in a greater degree of frustration~\cite{zheng;prb05}), it also exhibits AF LRO below \TN\ = 45~K [see Fig.~\ref{fig:structure}(b)]. It differs from the magnetic structure in CMO only in that the coupling of adjacent manganese layers stacked along the \textit{c} axis in NMO is ferromagnetic, in contrast to the antiferromagnetic coupling between layers in CMO~\cite{giot;prl07}. Moreover, significant short-range magnetic correlations survive well into the paramagnetic state in NMO~\cite{giot;prl07}. Surprisingly, no long-range structural phase transition accompanies the magnetic transition, implying that some other, unknown means must relieve the magnetic degeneracy. Recent work further suggests that the magnetic ground state is intrinsically inhomogeneous~\cite{zorko;nc14}, adding to the unusual nature of the magnetism in this compound. Finally, this system exhibits polymorphism~\cite{abaku;cm14} and significant structural disorder~\cite{orlan;prm18} that may influence the magnetostructural properties, although how this may happen remains an open question. Clarifying these issues in NMO continues to be an important objective in the study of $A$MnO$_2$ systems, while also presenting a valuable opportunity for a more general study of magnetic frustration, disorder, and their interplay in TMOs. This complements earlier work focused on magnetic pyrochlores and related materials~\cite{gardn;rmp10}.

Here, we discover that the magnetic phase transition in NMO occurs via an unconventional mechanism in which a nanoscale structural distortion supports macroscale magnetic order. Using atomic pair distribution function (PDF) analysis of neutron total scattering data~\cite{egami;b;utbp12}, we observe a short-range triclinic distortion in NMO correlated over a 2-nm length scale that exists well above \TN\ and grows significantly as the temperature decreases. Complementary magnetic PDF analysis~\cite{frand;aca14,frand;aca15,frand;prl16} reveals magnetic short-range order (SRO) persisting well above \TN, which likewise grows in magnitude and correlation length upon cooling until finally achieving the LRO state below \TN. Crucially, quantitative refinements confirm that the short-range triclinic distortion lifts the degeneracy of the isosceles triangular network on the nanoscale, thereby enabling the LRO state and enhancing the magnetic correlations above the transition.

While this short-range triclinic distortion is the primary actor, we find that other types of quenched structural disorder also play a role in determining the eventual ground state---specifically, Na vacancies in NMO, and antisite defects involving Cu atoms on the Mn site in CMO. Beyond a certain threshold, this type of disorder suppresses the long-range triclinic structural phase transition, resulting in the preservation of monoclinic symmetry in the average structure of NMO.  Taken together, these results describe an unusual mechanism for establishing magnetic LRO in the presence of frustration and a spatially inhomogeneous lattice, through which short-range correlations in the lattice sector promote long-range correlations in the magnetic sector.

\section{Materials and Methods}
Powder samples of \NMO\ and \CMO\ were prepared as described in Ref.~\cite{vecch;prb10,abaku;cm14}. Temperature-dependent time-of-flight neutron total scattering measurements were conducted using the Nanoscale-Ordered Materials Diffractometer (NOMAD) instrument~\cite{neufe;nimb12} at the Spallation Neutron Source (SNS) of Oak Ridge National Laboratory (ORNL) and the General Materials Diffractometer (GEM)~\cite{willi;pb97} at the ISIS neutron source of Rutherford Appleton Laboratory (RAL). A maximum momentum transfer of 35~\AA$^{-1}$ was used when generating the real-space PDF patterns. Analysis and modeling of the atomic PDF were performed with the PDFgui program~\cite{farro;jpcm07} and the DiffPy suite of programs for diffraction analysis~\cite{juhas;aca15}. The diffpy.mpdf package was used for the magnetic PDF analysis. Further details about the atomic and magnetic PDF refinements are provided in the Supplementary Information~\footnote{The Supplementary Information contains additional details about the atomic and magnetic pair distribution function analysis and the transmission electron microscopy characterization of \NMO\ and \CMO.}. Rietveld refinements were performed within the General Structure Analysis System (GSAS) software~\cite{toby;jac01} using data collected from the high-resolution backscattering banks of the GEM diffractometer.

\section{Results}
\textbf{\textit{Atomic PDF Analysis.}} 
We first present the atomic PDF analysis of NMO and CMO. The PDF, obtained from the sine Fourier transform of the reduced total scattering structure function $F(Q) = Q(S(Q)-1)$, provides a detailed account of the local atomic structure in real space, irrespective of the character of the long-range structure. Representative PDF patterns for both compounds at room temperature are shown in Fig.~\ref{fig:structure}(d). The expected monoclinic structural models explain the data well, vouching for the quality of the samples. Rietveld refinements of the Bragg intensities verified these results and allowed us to refine the average site occupancies. From this analysis, the actual compositions of our samples are Na$_{0.92}$MnO$_2$ and Cu(Mn$_{0.98}$Cu$_{0.02}$)O$_2$, in line with compositions reported in the literature~\cite{dally;nc18,dally;prb18}. For simplicity, we continue to label these samples as \NMO\ and \CMO. 

\begin{figure*}
	\includegraphics[width=140mm]{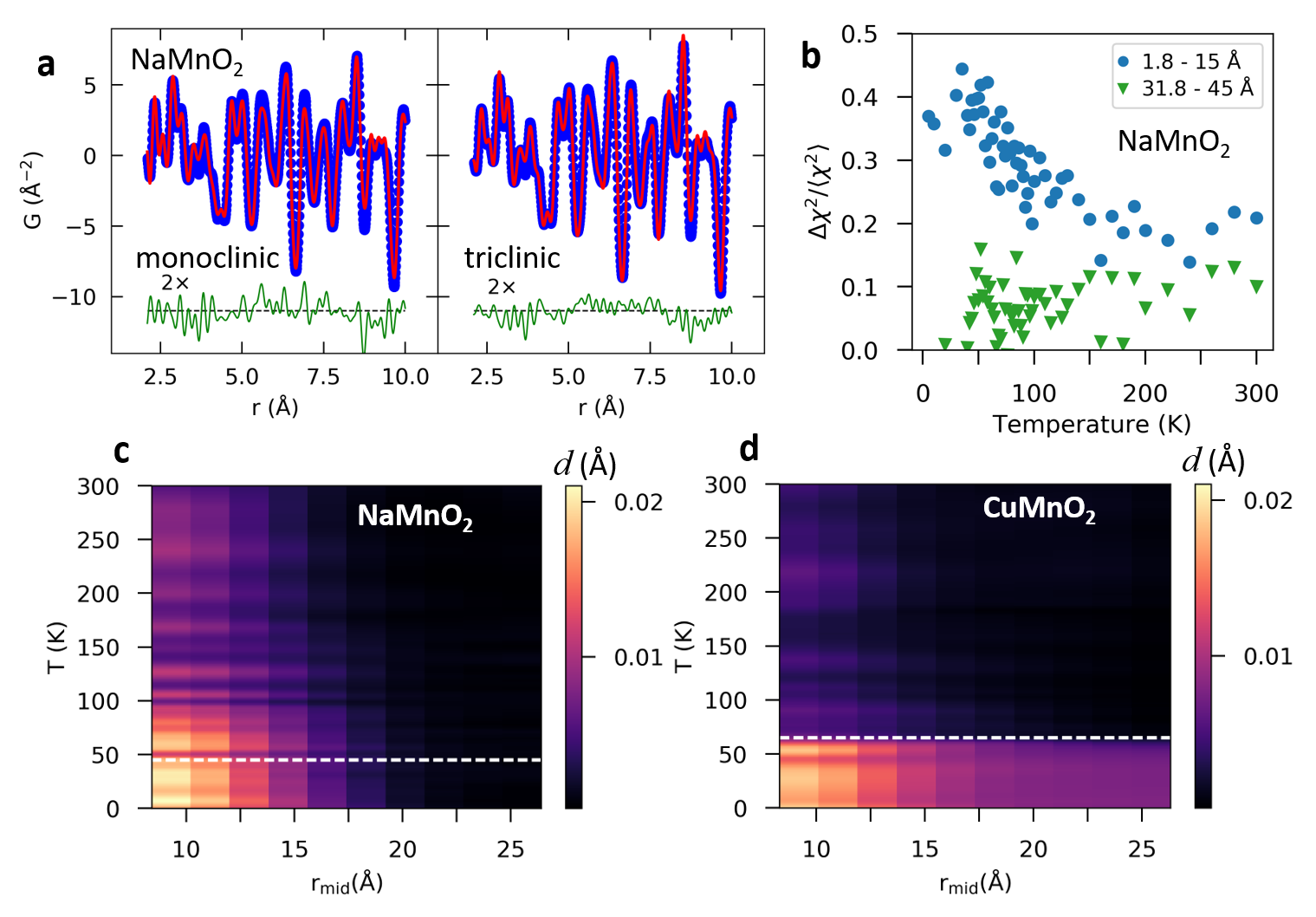}
	\caption{\label{fig:strucFitResults} \textbf{Results of atomic PDF fits for \CMO\ and \NMO.} (a) PDF refinements of the \NMO\ structure at 5~K using the monoclinic (left) and triclinic (right) models. The lower green curves are the fit residuals, multiplied by 2 for clarity. (b) Normalized difference in $\chi^2$ between the monoclinic and triclinic models of \NMO\ as a function of temperature for short (blue circles) and long (green triangles) fitting ranges. (c) Color map of the triclinic splitting in \NMO\ as a function of temperature and fitting range. The horizontal dashed line marks \TN. (d) Same as (c) but for \CMO.}
	
\end{figure*}

Intriguingly, the low-temperature PDF fits to the NMO data using the monoclinic model are worse at low $r$ ($<20$~\AA) than at high $r$, despite the fact that the average structure is known to be monoclinic at all temperatures. This implies that the local structure on such a short length scale deviates from the average monoclinic structure. Indeed, a triclinic structural model equivalent to that of CMO provides a significantly improved fit to the low-$r$ region at low temperature, as seen in Fig.~\ref{fig:strucFitResults}(a).
Inspections of the low-$r$ fit residuals (lower green curves) for the monoclinic and triclinic models of NMO at 5~K reveals that high-frequency features are consistently smaller for the triclinic model, resulting in a relative improvement in $\chi^2$ of 40\% over the monoclinic model. We therefore conclude that the ground state of NMO consists of regions of locally triclinic symmetry existing within the average monoclinic structure; hence, the structure is spatially inhomogeneous. We note that the PDF measurements cannot distinguish between static and dynamic distortions, so the temporal nature of this inhomogeneity is unclear. Importantly, this short-range triclinic distortion lifts the degeneracy of the triangular lattice locally, irrespective of the fact that the average crystal structure retains its monoclinic symmetry to the lowest temperature. The broad, lower-frequency signal seen in the fit residuals in Fig.~\ref{fig:strucFitResults}(a) arises from the magnetic contribution to the PDF, which we will discuss later.

Establishing the temperature dependence of this local triclinic distortion is crucial for understanding its connection to the magnetism in NMO. To this end, we refined both the monoclinic and triclinic structural models against the PDF data, which were collected on a dense temperature grid between 5 and 300~K. For each temperature $T$, fits were performed on a sliding data window ranging from [1.8~\AA\ - 15~\AA] to [31.8~\AA\ - 45~\AA] in 2-\AA\ steps. We then computed the normalized difference in $\chi^2$ between the monoclinic and triclinic models as $\Delta \chi^2 / \langle \chi^2 \rangle = \left(\chi^2_{\mathrm{mono}}-\chi^2_{\mathrm{tri}}\right)/\left(0.5\left[\chi^2_{\mathrm{mono}}+\chi^2_{\mathrm{tri}}\right]\right)$ for every refinement, quantifying the improvement offered by the triclinic model: the larger this normalized difference in $\chi^2$, the better the triclinic model relative to the monoclinic model.

In Fig.~\ref{fig:strucFitResults}(b), we plot $\Delta \chi^2 / \langle \chi^2 \rangle$ as a function of $T$ for the shortest and longest fitting ranges. The triclinic model performs significantly better than the monoclinic model for the short fitting range at low $T$. As $T$ increases, $\Delta \chi^2 / \langle \chi^2 \rangle$ steadily decreases until it plateaus at a nonzero value for temperatures above $\sim$150-200~K. For the longer fitting ranges, the triclinic model provides a markedly smaller relative improvement, with no pronounced changes as $T$ varies. These results confirm that the triclinic distortion in NMO has a short-ranged nature and is strongest at low temperature but may persist up to room temperature.

A more comprehensive view of the $T$- and $r$-dependence of this local distortion is given by the color map in Fig.~\ref{fig:strucFitResults}(c). The intensity scales with the triclinic splitting $d = r_2 - r_3$, which is responsible for lifting the local degeneracy of the isosceles triangles. Temperature appears on the vertical axis, and the horizontal axis displays the midpoint of the fitting range. To remove any spurious results arising from possible overfitting with the triclinic model, we scaled the $d$ values by the difference in $\chi^2$ between the monoclinic and triclinic models (see Supplementary Information for further details). The figure indicates that the greatest distortion ($d \sim 0.024$~\AA) occurs below $\sim$7~K and at low $r$. This distortion is roughly three times as large as that observed in the average structure of CMO ($d \sim 0.008 $~\AA). Averaging over longer distances reduces the distortion, evidenced by the weaker intensity on the right side of the color map. At low temperature, we estimate the correlation length of the local triclinic distortion to be about 20~\AA. With increasing temperature, the magnitude of the distortion decreases, although it remains nonzero even at room temperature. 

We display the equivalent results for CMO in Fig.~\ref{fig:strucFitResults}(d). Because of the long-range structural phase transition at 65~K, the triclinic splitting remains nonzero over the entire real-space range below this temperature, albeit with somewhat enhanced splitting for short distances, suggesting incomplete ordering. Above 65~K, short-range triclinicity remains, but with a smaller magnitude and shorter real-space extent than in NMO. Overall, we observe that NMO displays significantly more spatial inhomogeneity in the form of local triclinic distortions than does CMO.

Analysis of the local octahedral environment reveals that no dramatic changes in the MnO$_6$ octahedra occur in either compound as the temperature is varied. The splitting of the equatorial Mn-O bonds is commensurate with the overall triclinic splitting of the $a$ and $b$ lattice parameters ($r_2$ and $r_3$ in Fig.~\ref{fig:structure}). The PDF data at room temperature reveal that the magnitude of the Jahn-Teller (JT) distortion, defined as the ratio of the apical to equatorial Mn-O bonds ($r_{ap}/r_{eq}$) in the MnO$_6$ octahedra, is smaller in \CMO\ (1.173; $r_{ap}=2.258(1) \mathrm{\AA}$, $r_{eq}=1.925(1) \mathrm{\AA}$) than in \NMO\ (1.240; $r_{ap}=2.390(1) \mathrm{\AA}$, $r_{eq}=1.928(1) \mathrm{\AA}$). The JT distortion in \CMO\ shows a very slight decrease of 0.1\% as the temperature is lowered from room temperature to 5 K. In \NMO, the JT distortion decreases by 0.6\% over that temperature range. In the triclinic state, the equatorial bonds become slightly unequal, so we take their average when computing the distortion magnitude.

\begin{figure*}
	\includegraphics[width=160mm]{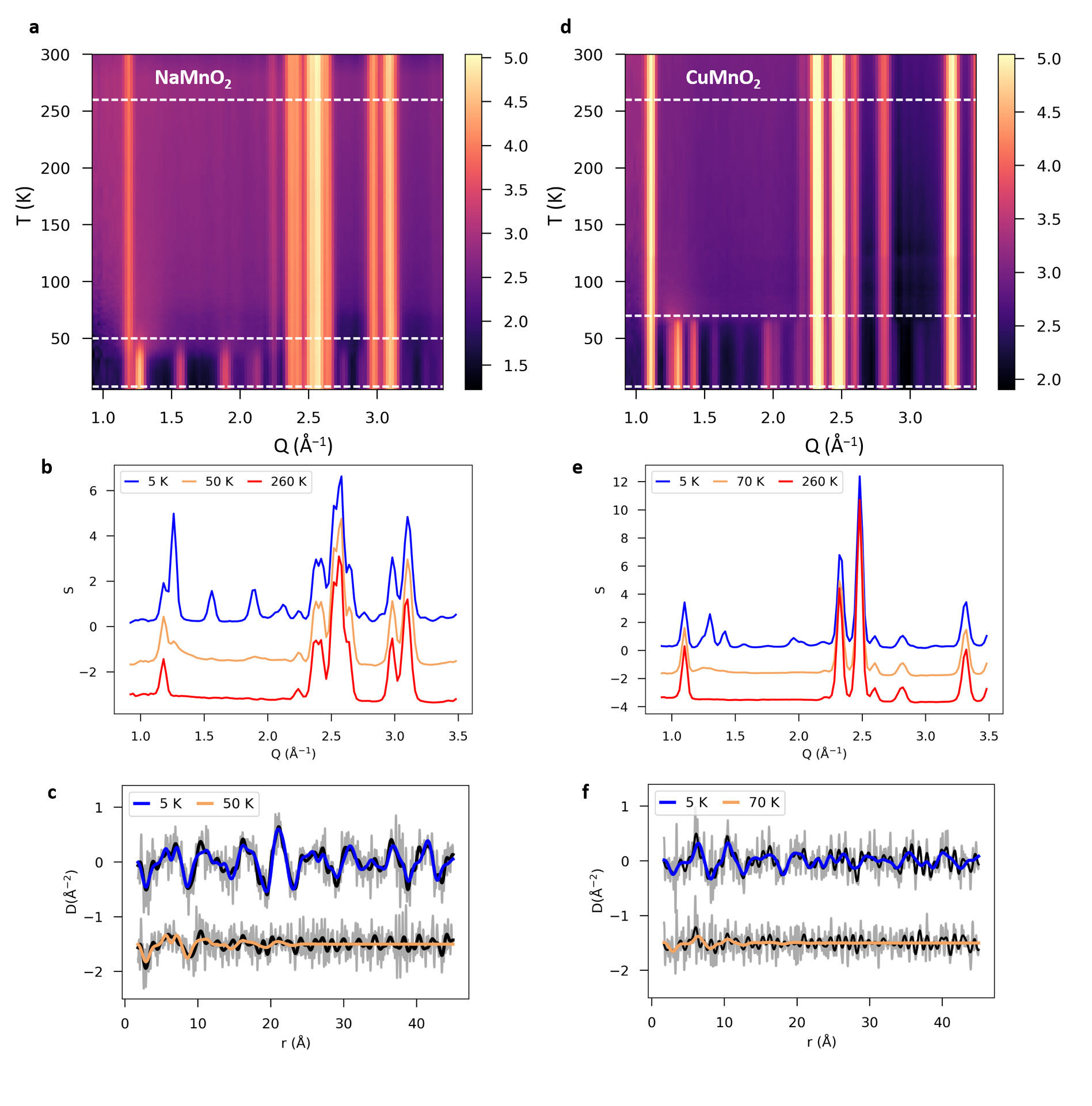}
	\caption{\label{fig:SQcmap} \textbf{Magnetic scattering and magnetic PDF analysis of \CMO\ and \NMO.} (a) Color map of the scattered intensity $S(Q)$ of \NMO\ with temperature on the vertical axis and $Q$ on the horizontal axis. (b) $S(Q)$ for \NMO\ at 5~K (blue curve), 50~K (orange curve), and 260~K (red curve), indicated by horizontal dashed lines in panel (a). (c) Corresponding mPDF data for \NMO\ at 5~K and 50~K. The 50~K data is offset along the y-axis by -1.5~\AA$^{-2}$ for clarity. The black curves and thin gray curves represent the Fourier filtered and unfiltered data, respectively. The blue and orange curves represent the mPDF fits for 5 and 50~K, respectively. (d-f) Equivalent results for \CMO. The intensity scale for (d) was chosen such that the strongest magnetic Bragg peaks for CMO and NMO have equal contrast. The magnetic model in Ref.~\onlinecite{vecch;prb10} was used for the fits in (f).}
\end{figure*}

\textbf{\textit{Magnetic PDF Analysis.}}
Having investigated the local atomic structures of NMO and CMO, we now turn to their magnetic structures. Fig.~\ref{fig:SQcmap}(a) displays a color map of the total scattering structure function $S(Q)$ for NMO, with temperature on the vertical axis and $Q$ on the horizontal axis. 
The nuclear Bragg peaks are clearly visible as the intense, temperature-independent features. Below 45~K, several well-defined magnetic Bragg peaks exist between the nuclear peaks, consistent with the expected long-range AF order. Some diffuse scattering remains visible below 45~K, likely due to the presence of mixed paramagnetic and AF phases in the vicinity of the transition. As the temperature increases, the magnetic LRO melts, causing the sharp magnetic Bragg peaks to dissipate into diffuse scattering that persists up to the highest temperatures measured and extends over a broad $Q$-range. Importantly, careful inspection of this diffuse scattering reveals that its $Q$-dependence remains non-uniform well above the transition temperature, indicative of persistent short-range magnetic correlations. To illustrate this further, we display in Fig.~\ref{fig:SQcmap}(b) cuts of $S(Q)$ at 5~K, 50~K, and 260~K. The strong and sharp magnetic Bragg peaks at 5~K give way to a weaker and more diffuse signal at 50~K, yet distinct features are still observable in the scattering pattern. At 260~K, any structure in the diffuse scattering is much less well defined.

Additional insight can be gained by analyzing the magnetic correlations directly in real space using the magnetic PDF (mPDF) method, which is sensitive to local magnetic structure even in the absence of long-range magnetic order~\cite{frand;prl16,frand;prm17}. Experimentally, the mPDF signal is contained in the fit residual produced by subtracting the best-fit atomic PDF from the total PDF signal; additional details appear in the Supplementary Information. Representative mPDF patterns for NMO at 5~K and 50~K are displayed in Fig.~\ref{fig:SQcmap}(c). These patterns correspond to the long-wavelength oscillatory signal in the atomic PDF fit residuals shown in Fig.~\ref{fig:strucFitResults}(a). The thin gray curves show the unfiltered mPDF signal mixed in with high-frequency noise, which is propagated through the Fourier transform due to the large value of $Q_{\mathrm{max}}$ (chosen to optimize the atomic PDF data). Applying a Fourier filter to remove all frequencies above 6.9~\AA$^{-1}$, where the squared magnetic form factor drops below 1\% its maximal value, results in the thick black curves in Fig.~\ref{fig:SQcmap}(c). This provides a cleaner signal while preserving the contribution from the magnetic scattering. The overlaid colored curves are fits that will be described subsequently. Simple inspection of the mPDF data at 5~K reveals a well-defined signal over the full $r$-range displayed, indicative of magnetic LRO. The alternating negative and positive peaks reflect the AF spin configuration~\cite{frand;aca14}. At 50~K, however, the signal is reduced in amplitude and fully suppressed in real space beyond about 20~\AA, pointing to short-range magnetic correlations with a length scale of $\sim$2~nm. 

To investigate the magnetic correlations more quantitatively, we modeled the mPDF using the known LRO AF structure published in Ref.~\onlinecite{giot;prl07}. Our model included as free parameters a scale factor (proportional to the square of the locally ordered moment), a real-space correlation length $\xi$, and two polar angles defining the spin direction. The mPDF fits shown in Fig.~\ref{fig:SQcmap}(c) for 5 and 50~K agree well with the data. At 5~K, the correlation length is limited by the instrumental resolution, and the refined spin direction matches previously published results~\cite{giot;prl07}. Earlier work reported intrinsic inhomogeneities in the magnetic ground state in the form of nanometer-sized bubbles possessing reversed sublattice magnetization relative to the surrounding matrix of Mn spins~\cite{zorko;nc14}. The presence of these bubbles would cause an additional damping in the mPDF signal, which we do not observe. Therefore, any such low-temperature magnetic inhomogeneity, if present, must exist on a length scale of $\sim$5~nm or greater, beyond the sensitivity of the mPDF data. We note that the incommensurability of the magnetic order reported in Ref.~\onlinecite{dally;prb18} would result in a real-space modulation outside the accessible data window of these measurements. At 50~K, the fit yields a correlation length of $\sim$15~\AA\ with a nearly unchanged spin direction. Interestingly, the fit can be improved by about 5\% if only a single two-dimensional (2D) layer is considered and by 8\% if just a one-dimensional (1D) chain of nearest-neighbor spins is used, suggesting that the short-range magnetic correlations develop a lower-dimensional character above \TN. This agrees with earlier inelastic neutron scattering analysis~\cite{stock;prl09,dally;nc18,dally;prb18}.

\begin{figure}
	\includegraphics[width=70mm]{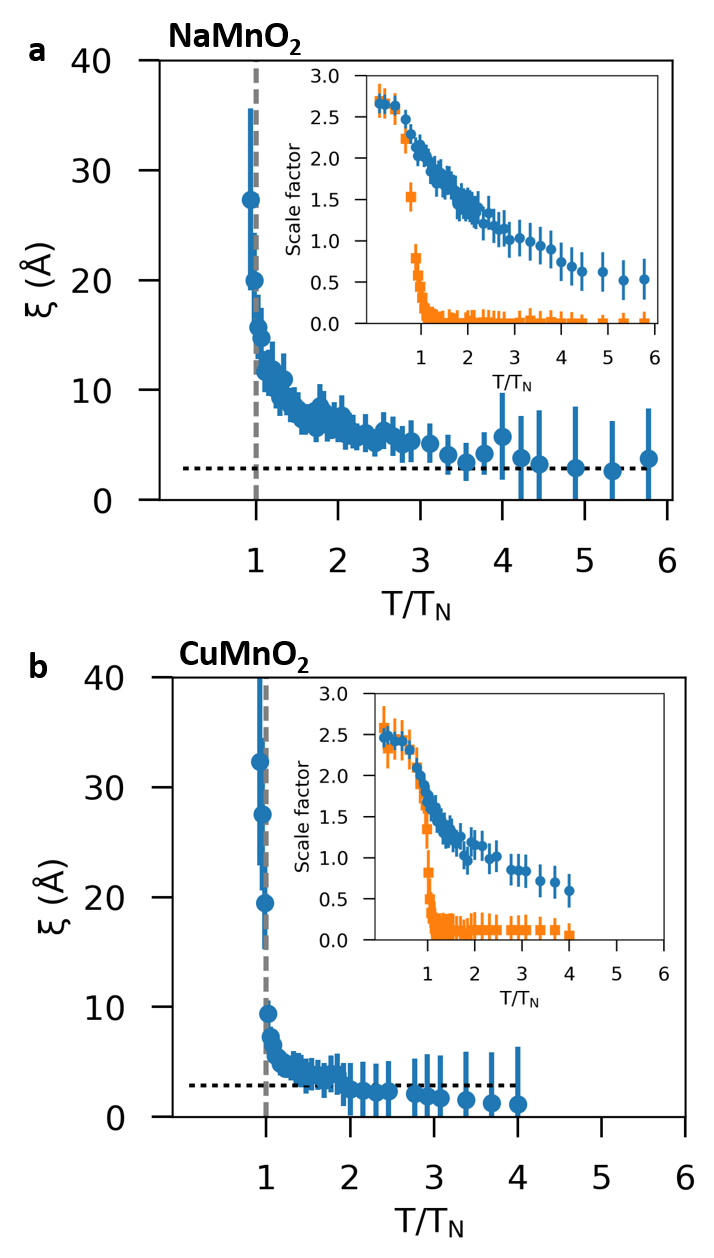}
	\caption{\label{fig:mPDF-both} \textbf{Temperature evolution of magnetic correlations in \CMO\ and \NMO.} (a) Correlation length of the short-range antiferromagnetic order in \NMO\ determined from mPDF refinements using a three-dimensional model, displayed as a function of dimensionless temperature $T/T_{\mathrm{N}}$. Below the transition (indicated by the vertical dashed line), the correlation length is limited by the instrumental resolution. The horizontal dashed line indicates the nearest-neighbor Mn-Mn distance. Inset: Refined mPDF scale factor as a function of dimensionless temperature for a fitting range of 1.8 to 15~\AA\ (blue circles) and 30 to 45~\AA\ (orange squares). (b) Equivalent results for \CMO.}
	
\end{figure}

In Fig.~\ref{fig:mPDF-both}(a), we display the refined magnetic correlation length $\xi$ plotted against the dimensionless temperature $T/T_{\mathrm{N}}$. Beginning slightly above $3T_{\mathrm{N}}$, $\xi$ is approximately equal to the Mn-Mn nearest-neighbor (NN) distance (illustrated by the horizontal dashed line). Large error bars above $4T_{\mathrm{N}}$ limit our ability to specify $\xi$ at these temperatures.
Below $3T_{\mathrm{N}}$, the correlation length steadily increases as the temperature decreases, until it reaches its resolution-limited value below the ordering temperature of $\sim$45~K, marked by the vertical dashed line. The inset of Fig.~\ref{fig:mPDF-both}(a) illustrates the temperature dependence of the mPDF scale factor for refinements over 1.8 - 15~\AA\ and 31.8 - 45~\AA. For the longer fitting range, the scale factor is zero at high temperature and shows a sharp increase for $T < T_{\mathrm{N}}$, indicating a rapid onset of long-range magnetic order. In contrast, the short-range scale factor shows a much more gradual decrease as the temperature increases, remaining nonzero to 300~K, the highest measured temperature. This quantifies the persistence of significant short-range magnetic correlations at temperatures far above the long-range transition, as expected from our previous qualitative analysis of the scattering data (Fig.~\ref{fig:SQcmap}). We note that the correlation length and short-range scale factor track proportionally with the amplitude of the local lattice distortion, as seen by comparing the evolution of the local atomic structure in Fig.~\ref{fig:strucFitResults}(b,c) with that of the magnetic correlations in Fig.~\ref{fig:mPDF-both}a, suggesting that they are closely connected.

The corresponding results for CMO are presented in Fig.~\ref{fig:SQcmap}(d-f) and Fig.~\ref{fig:mPDF-both}(b). Qualitatively, CMO shows similar magnetic behavior as NMO, with short-range correlations surviving into the paramagnetic phase above $T_{\mathrm{N}} = 65$~K. Quantitatively, however, we find that these short-range correlations in CMO are weaker and more limited in spatial extent than for NMO. As seen in Fig.~\ref{fig:mPDF-both}, the correlation length and short-fitting-range scale factor decrease above \TN\ significantly more rapidly in CMO than in NMO.

\section{Discussion}

The key experimental findings obtained from the PDF and mPDF analyses are: (1) NMO exhibits a short-range triclinic distortion correlated over $\sim$2~nm, lifting the magnetic degeneracy locally; (2) this distortion grows in amplitude upon cooling; (3) short-range AF correlations are present in NMO at least up to 300~K and grow in strength and correlation length upon cooling, tracking with the structural distortion; and (4) CMO exhibits a smaller local triclinic distortion and weaker AF correlations than does NMO.

Several insights can be gleaned from these results. First, they clarify the structural ground state of NMO, demonstrating that the average structure remains monoclinic, while the instantaneous local structure is triclinically distorted. Second, the observation that the triclinic distortion and magnetic SRO both grow upon cooling with a very similar temperature dependence indicates that these two effects are closely related. The relatively weaker local structural distortion and AF correlations in the paramagnetic phase of CMO compared to NMO further supports this. Third, the fact that the short-range structural distortion relieves the local magnetic degeneracies naturally suggests that it promotes the growth of the AF magnetic correlations. Intriguingly, the AF correlation length increases upon cooling until matching the $\sim$2~nm length scale of the local triclinic distortion around \TN, at which point magnetic LRO occurs. This may indicate that magnetic coherence across neighboring triclinic regions triggers the transition. Together, these findings provide a cohesive picture to understand the magnetic transition in NMO, in which short-range structural distortions lift the magnetic degeneracy on the nanoscale and eventually enable long-range magnetic order. In a more general context, NMO appears to be a rare example of a geometrically frustrated magnet exhibiting long-range magnetic order coupled to a short-range lattice distortion. Separately, the local triclinic distortion observed in NMO by PDF may explain various unusual experimental results in the literature, including the subtle anisotropic broadening of the low-temperature Bragg peaks seen in previous diffraction work~\cite{giot;prl07}, the nanoscale inhomogeneities of the ground state suggested by muon spin relaxation and nuclear magnetic resonance experiments~\cite{zorko;nc14,zorko;sr15}, and the emergent magnetoelectric response~\cite{orlan;prm18}.

Given the fundamentally similar structural and magnetic ingredients of NMO and CMO, a natural question to ask is why NMO accomplishes magnetic LRO via a short-range structural distortion, while CMO does the same with a more conventional long-range phase transition. One possibility arises from earlier ab-initio calculations, which predicted a larger energy gain resulting from the triclinic structure in the AF state for CMO than NMO~\cite{zorko;sr15}. However, somewhat surprisingly, the PDF analysis demonstrates that NMO is actually more triclinic on a nanometer length scale than the average CMO triclinic distortion, counter to the predictions. Another possibility would be the presence of stronger magnetoelastic coupling in CMO than in NMO, yet this is also unlikely to be the explanation, as calculations have shown that the magnetoelastic energy scale in both materials is 2-3 orders of magnitude weaker than the magnetic exchange interactions~\cite{zorko;sr15}.

Instead, we suggest that structural disorder plays a decisive role. In these complex systems, multiple types of disorder are present. First, the local triclinic distortion discussed previously is present in both systems and plays an active role in promoting AF correlations by lifting the magnetic degeneracy locally. Second, $A$-site occupancy disorder in the form of Na vacancies and excess Cu on the Mn site exist in NMO and CMO, respectively. Third, planar defects due to oxygen-layer gliding exist in both materials. The experimental evidence here and elsewhere indicates that NMO exhibits significantly more structural disorder than CMO. This is evidenced by the more pronounced distortion of the local structure in NMO than CMO, the greater percentage of site occupancy disorder (8\% Na vacancies versus 2\% excess Cu from antisite defects), and a higher density of planar defects (see Supplementary Information)~\cite{abaku;cm14,cleme;jes15,orlan;prm18}. Note that the Na vacancies and Cu impurities allow for conversion of Mn$^{3+}$ ($S=2$) to Mn$^{4+}$ ($S=3/2$)~\cite{li;nm14}, which could also create magnetic disorder. It is tempting to relate such mixed-valency mediated disorder to Zener-type carriers~\cite{zener;pr51a,zener;pr51b}, which may cause local inhomogeneities in the spin structure when entering bound states~\cite{degen;pr60}. Verifying this scenario is beyond the resolving capabilities of our experimental probes.

These afore-mentioned factors of quenched disorder~\cite{burgy;prl01} contribute to significant exchange disorder, i.e. randomness in the magnetic exchange interactions between nearby spins. Theoretically, the NN Heisenberg model in the presence of arbitrarily weak magnetoelastic coupling predicts a critical level of exchange disorder. Below this level, a frustrated system enters the LRO N\'eel state through a concomitant lowering of its average lattice symmetry; above the critical level, the N\'eel state arises without any long-range structural distortion~\cite{saund;prl07,saund;prb08}. Our results suggest that CMO rests below the critical level of disorder and NMO above it, although differing degrees of sensitivity to perturbations between the two systems may also contribute~\cite{moess;cjp01}. Considering also that NMO is more frustrated than CMO and likely exhibits stronger electron-lattice coupling via an enhanced Jahn-Teller (JT) distortion~\cite{giot;prl07,vecch;prb10}, the larger exchange disorder may lock the spin order within the local triclinicity and thus help NMO exit degeneracy without a long-range, symmetry-lowering phase transition as in the less frustrated and weaker JT-distorted CMO. 

These arguments are in line with the observation that CuMn$_{1-x}$Cu$_x$O$_2$ displays a long-range triclinic distortion for $x < 7\%$ (which includes the current CMO sample with $x = 2\%$), but retains the high-symmetry monoclinic average structure for more disorderd compounds with $x \ge 7\%$~\cite{garle;prb11,poien;cm11}. In this picture, compounds with larger values of $x$ possess a level of exchange disorder that exceeds the critical level, preventing the long-range structural transition to the triclinic phase, as in the case of \NMO. Additionally, the magnetic structure in CuMn$_{1-x}$Cu$_x$O$_2$ exhibits antiferromagnetic coupling between layers for low values of $x$ in which the triclinic phase is observed, but changes to ferromagnetic coupling between layers for higher $x$ where the long-range triclinic transition is suppressed~\cite{terad;prb11,garle;prb11}. \NMO\ also shows ferromagnetic coupling between the layers, further supporting the analogy between \NMO\ and heavily Cu-doped \CMO. Whether the CuMn$_{1-x}$Cu$_x$O$_2$ system with $x \ge 7\%$ also shows a local triclinic distortion and significant short-range magnetic correlations at high temperature like NMO would be an informative follow-up study. More generally speaking, we note that the importance of local structural disorder revealed here supports similar recent findings in the canonical pyrochlore Yb$_2$Ti$_2$O$_7$~\cite{bowma;nc19}.

In summary, the combined atomic and magnetic PDF analysis of NMO reveals an unusual mechanism of relieving geometrical frustration in which a short-range lattice distortion results in long-range magnetic order. Comparison with CMO in the paramagnetic phase further establishes the close relationship between the short-range structural and antiferromagnetic correlations, while also highlighting the role that structural disorder such as Na vacancies and Cu antisite defects plays in ground state selection for frustrated systems. More generally, this work illuminates the cooperative intertwining of the local atomic and magnetic structures that can occur when spatial inhomogeneity meets geometrical frustration. We suggest that similar physics may be at play in numerous other TMO systems hosting geometrically frustrated magnetism. Finally, the nanoscale magnetic degeneracy lifting that occurs above \TN\ shares similarities with the nanoscale \textit{orbital} degeneracy lifting observed in other systems, such as the manganites~\cite{qiu;prl05}, iron-chalcogenides~\cite{frand;prb19,koch;prb19}, and iridates~\cite{bozin;sr14,bozin;nc19}. Together, these findings contribute to an emerging understanding of the importance of spontaneous local degeneracy lifting that appears above long-range, symmetry-broken ground states in complex materials.

\vspace{2mm}

\begin{acknowledgments}
	
	\textit{Acknowledgments.} This research used resources at the Spallation Neutron Source, a DOE Office of Science User Facility operated by the Oak Ridge National Laboratory. ORNL is managed by UT-Battelle, LLC, under Contract No. DE-AC0500OR22725 for the U.S. Department of Energy. B.A.F. acknowledges support from Brigham Young University. Work at Brookhaven National Laboratory was supported by US DOE, Office of Science, Office of Basic Energy Sciences (DOE-BES) under Contract No. DE-SC0012704. This work has been partially supported by U.S. DOE Grant No. DE-FG02-13ER41967. We thank the Science and Technology Facilities Council (STFC) for the provision of neutron beam time at ISIS Facility. We acknowledge Dr. J\"org Neuefeind for assistance with the measurements at the NOMAD beamline and Dr. Dave Keen for assistance on the GEM beamline. Access to the facilities of the Integrated Infrastructure Initiative ESTEEM2 (EU 7th FP program with reference number 312483) is gratefully acknowledged. We are indebted to Prof. Artem M. Abakumov for the TEM-based characterizations at EMAT. AL thanks the Fulbright Foundation - Greece.
	
\end{acknowledgments}

%

\end{document}


\title{
Supplementary Information: Nanoscale Degeneracy Lifting in a Geometrically Frustrated Antiferromagnet
}

\author{Benjamin A. Frandsen}
\email{benfrandsen@byu.edu}
\affiliation{ %
	Department of Physics and Astronomy, Brigham Young University, Provo, Utah 84602, USA.
} %

\author{Emil S. Bozin}
\email{bozin@bnl.gov}
\affiliation{%
	Condensed Matter Physics and Materials Science Division, Brookhaven National Laboratory, Upton, NY 11973, USA.
}%

\author{Eleni Aza}
\affiliation{
	Institute of Electronic Structure and Laser, Foundation for Research and Technology---Hellas, Vassilika Vouton, 71110 Heraklion, Greece.
}
\affiliation{
	Department of Materials Science and Engineering, University of Ioannina, 451 10 Ioannina, Greece.
}

\author{Antonio Fern\'andez Mart\'inez}
\affiliation{
	Institute of Electronic Structure and Laser, Foundation for Research and Technology---Hellas, Vassilika Vouton, 71110 Heraklion, Greece.
}

\author{Mikhail Feygenson}
\affiliation{
	Neutron Scattering Division, Oak Ridge National Laboratory, Oak Ridge, Tennessee 37831, USA.
}%
\affiliation{
	J\"ulich Centre of Neutron Science, Forschungszentrum J\"ulich, 52428, J\"ulich Germany.
}%

\author{Katharine Page}
\affiliation{
	Neutron Scattering Division, Oak Ridge National Laboratory, Oak Ridge, Tennessee 37831, USA.
}%
\affiliation{
	Materials Science and Engineering Department, University of Tennessee, Knoxville, Tennessee 37996, USA.
}%

\author{Alexandros Lappas}
\email{lappas@iesl.forth.gr}
\affiliation{
	Institute of Electronic Structure and Laser, Foundation for Research and Technology---Hellas, Vassilika Vouton, 71110 Heraklion, Greece.
}

\maketitle

This Supplementary Material contains additional details about the atomic and magnetic pair distribution function analysis and the transmission electron microscopy characterization of \NMO\ and \CMO.

Atomic PDF fits were performed as a function of temperature and fitting range for \CMO\ and \NMO\ using both the monoclinic (space group C2/m) and triclinic (space group P$\overline{1}$) structural models. In each case, all symmetry-allowed structural parameters for the given space group were refined, including the lattice vectors and angles, the displacive degrees of freedom for the oxygen atoms, and the anisotropic displacement parameters corresponding to thermal vibrations. The linear correlation parameter to correct for correlated atomic motion at low $r$ was also refined as a temperature-dependent parameter for all fits containing data with $r \le 6$~\AA, following established PDF analysis protocols. The instrumental parameters $Q_{damp}$ and $Q_{broad}$ were determined from independent fits performed at high temperature and were fixed for all other refinements. For NOMAD, we used $Q_{damp} = 0.0354~\mathrm{\AA}^{-1}$ and $Q_{broad} = 0.0415~\mathrm{\AA}^{-1}$, while for GEM the values were $Q_{damp} = 0.0329~\mathrm{\AA}^{-1}$ and $Q_{broad} = 0.0439~\mathrm{\AA}^{-1}$. An overall scale factor was also refined. In Fig.~2(c,d) in the Main Text, the values of the triclinic splitting $d = r_2 - r_3$ determined from refinements at each $(r_{\mathrm{mid}},T)$ point were scaled by the dimensionless quantity $\Delta \chi^2 (r_{\mathrm{mid}}, T)/ \Delta \chi^2 _{\mathrm{max}}$, where $\Delta \chi^2  = \chi^2_{\mathrm{mono}}-\chi^2_{\mathrm{tri}}$ and $\Delta \chi^2 _{\mathrm{max}}$ is the maximal difference in $\chi^2$, observed at the lowest temperature and shortest fitting range. This scaling prevents the possibility of an exaggerated triclinic splitting due to overfitting with the triclinic model if it provides no significant benefit over the monoclinic model.

The magnetic PDF data were obtained from the NOMAD total PDF data using the following procedure: 1) We first refined the atomic structure and subtracted the best-fit atomic PDF from the experimental PDF at each temperature. This produces a fit residual containing the mPDF and any errors arising from deficiencies in the structural model or imperfections in the PDF data arising from noise or incomplete background subtraction. 2) We then subtracted the fit residual at 280~K from those at all lower temperatures, thereby removing temperature-independent features from the data such as artifacts from the background, helping to isolate the mPDF further. Magnetic PDF refinements were performed against these data sets. Similar subtraction of data sets is routinely performed to isolate magnetic scattering in conventional magnetic neutron diffraction. 3) For purposes of data visualization, we also performed a Fourier-filter operation on the mPDF signal to remove all frequencies above 6.9~\AA$^{-1}$, corresponding to the value of $Q$ at which the square of the magnetic form factor for Mn$^{3+}$ is approximately 1\% its maximal value. The resulting filtered mPDF signals are illustrated by the thick, black curves in Fig.~3(c,f) of the Main Text.

\begin{figure}
	\includegraphics[width=60mm]{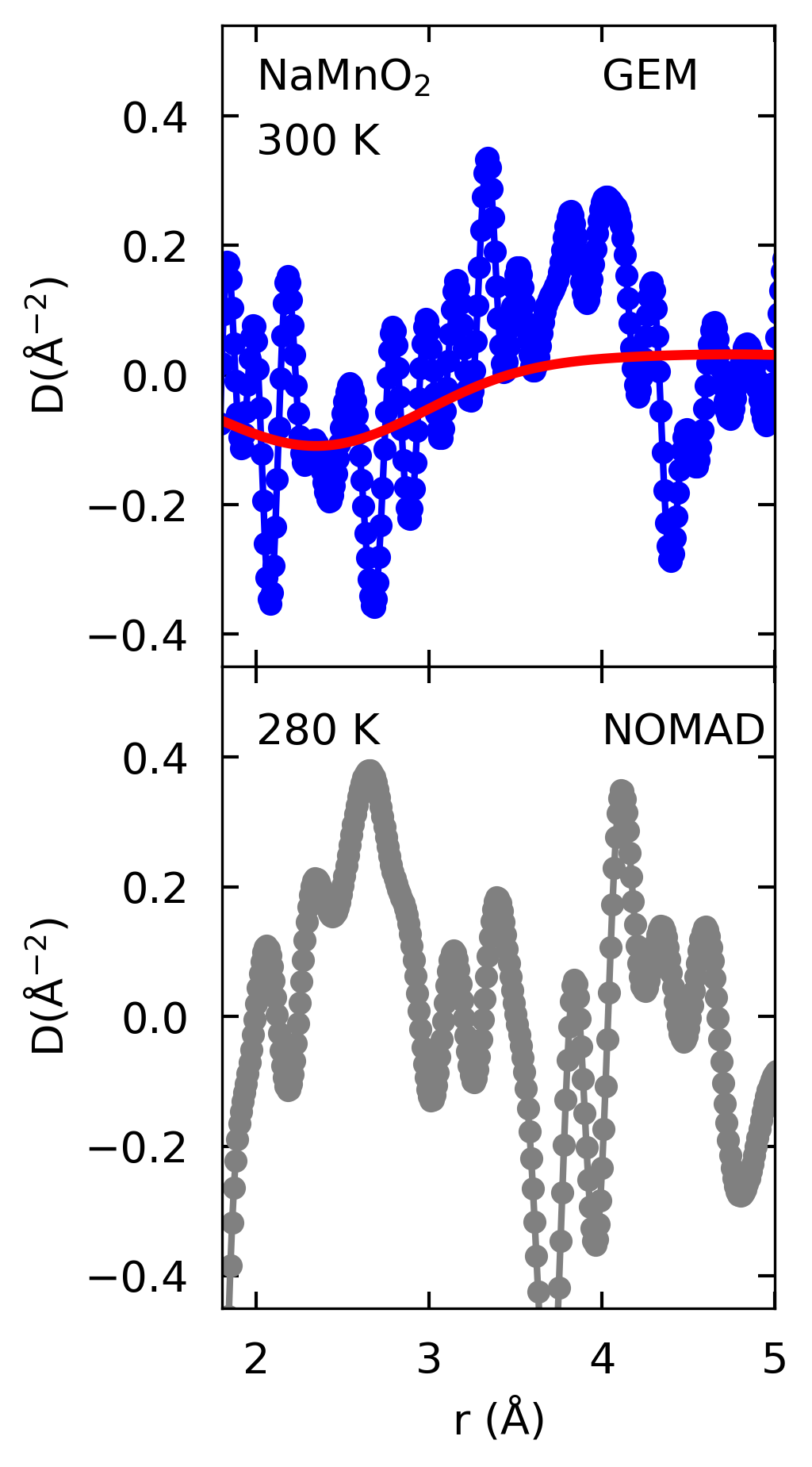}
	\caption{\label{fig:exampleFit} Top: Atomic PDF fit residual for \NMO\ measured on GEM at 300~K and modeled with the triclinic structure (solid blue circles), along with the best-fit mPDF pattern (solid red curve). The downturn between 2 and 3~\AA\ is due to antiferromagnetic nearest-neighbor correlations. Bottom: Equivalent atomic PDF fit residual for \NMO\ measured on NOMAD at 280~K.}
	
\end{figure}

We note that Step 2 of this procedure causes any residual magnetic PDF present in the data at 280 K to be subtracted out from all lower temperatures. Therefore, the mPDF patterns at lower temperatures represent the local magnetic configuration relative to that at 280 K. The measurements performed at GEM provide an important independent check on the validity of this three-step procedure. Collecting equivalent data on two different instruments helps eliminate potentially spurious signals in the data contributed by the instruments themselves. Because the background in the GEM data was lower, the subtraction of the high-temperature residual was unnecessary. After applying Step 1 to the GEM data, we determined that the only observable component of the mPDF near room temperature was a small negative peak centered on the Mn-Mn nearest neighbor distance ($\sim$2.86~\AA) corresponding to antiferromagnetic correlations. An enlarged view of the atomic PDF fit residual for \NMO\ at 300~K measured at GEM is shown by the blue filled circles in the top panel of Fig.~S\ref{fig:exampleFit}.
The solid red curve shows the best-fit mPDF using a one-dimensional model consisting of a single chain of AF-coupled Mn moments along the Mn-Mn nearest-neighbor direction in the lattice. This model captures the downturn between 2 and 3~\AA\ nicely, corresponding to the Mn-Mn nearest neighbor antiferromagnetic correlations. This feature grows in magnitude as the temperature is lowered, as expected from the development of stronger magnetic correlations. The loss of this small antiferromagnetic nearest-neighbor feature from the NOMAD data does not affect the interpretation of the results, and the high-temperature scale factor determined on GEM was incorporated into the scale factors displayed in Fig. 4 of the Main Text. The positive feature seen around 4~\AA\ in the top panel of Fig.~S\ref{fig:exampleFit} cannot be magnetic in origin, since it is completely independent of temperature and there are no Mn-Mn pairs at that distance. For comparison, the bottom panel of Fig.~S\ref{fig:exampleFit} shows the atomic PDF fit residual for data collected on NOMAD at 280~K. This is the fit residual that was subtracted from all lower-temperature fits to the NOMAD data. Due to the larger background and greater overall magnitude of the fit residual for the NOMAD data compared to the GEM data, the AF correlations around the Mn-Mn nearest-neighbor distance are less evident, reinforcing the value of independent measurements on GEM. 

\begin{figure*}
	\includegraphics[width=150mm]{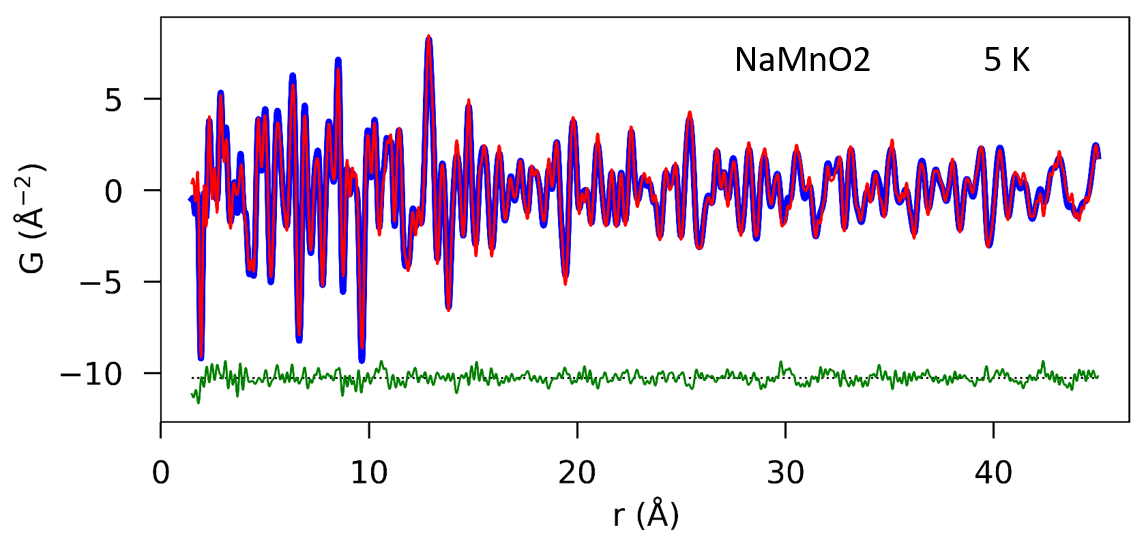}
	\caption{\label{fig:combinedFit} Combined atomic and magnetic PDF fit (red curve) overlaid on the experimental PDF data (blue curve) collected from \NMO\ at 5~K. The difference curve is shown in green, offset vertically for clarity.}
	
\end{figure*}

We refined the following four parameters in our mPDF fits: A scale factor, which is proportional to the square of the locally ordered magnetic moment; an exponential damping parameter encoding the finite magnetic correlation length, which was applied to the calculated mPDF in conjunction with the ``$Q_{damp}$'' Gaussian envelope determined from atomic PDF refinements~\cite{farro;jpcm07}; and polar angles $\theta$ and $\phi$ that set the direction of the magnetic moments. The relative orientation of neighboring moments was determined by the propagation vectors published previously: $\mathbf{k} = (0,1/2,1/2)$ in the triclinic setting for \CMO\ and $\mathbf{k} = (1/2,1/2,0)$ in the monoclinic setting for \NMO~\cite{vecch;prb10,giot;prl07}. For both compounds, the spin direction refined at low temperature agreed with the published spin directions to within the uncertainty of the refinements. In Fig.~S\ref{fig:combinedFit}, we display the end result of this fitting procedure, including both the atomic and magnetic PDF contributions, for \NMO\ at 5~K over a long $r$-range.

\begin{figure}
	\includegraphics[width=80mm]{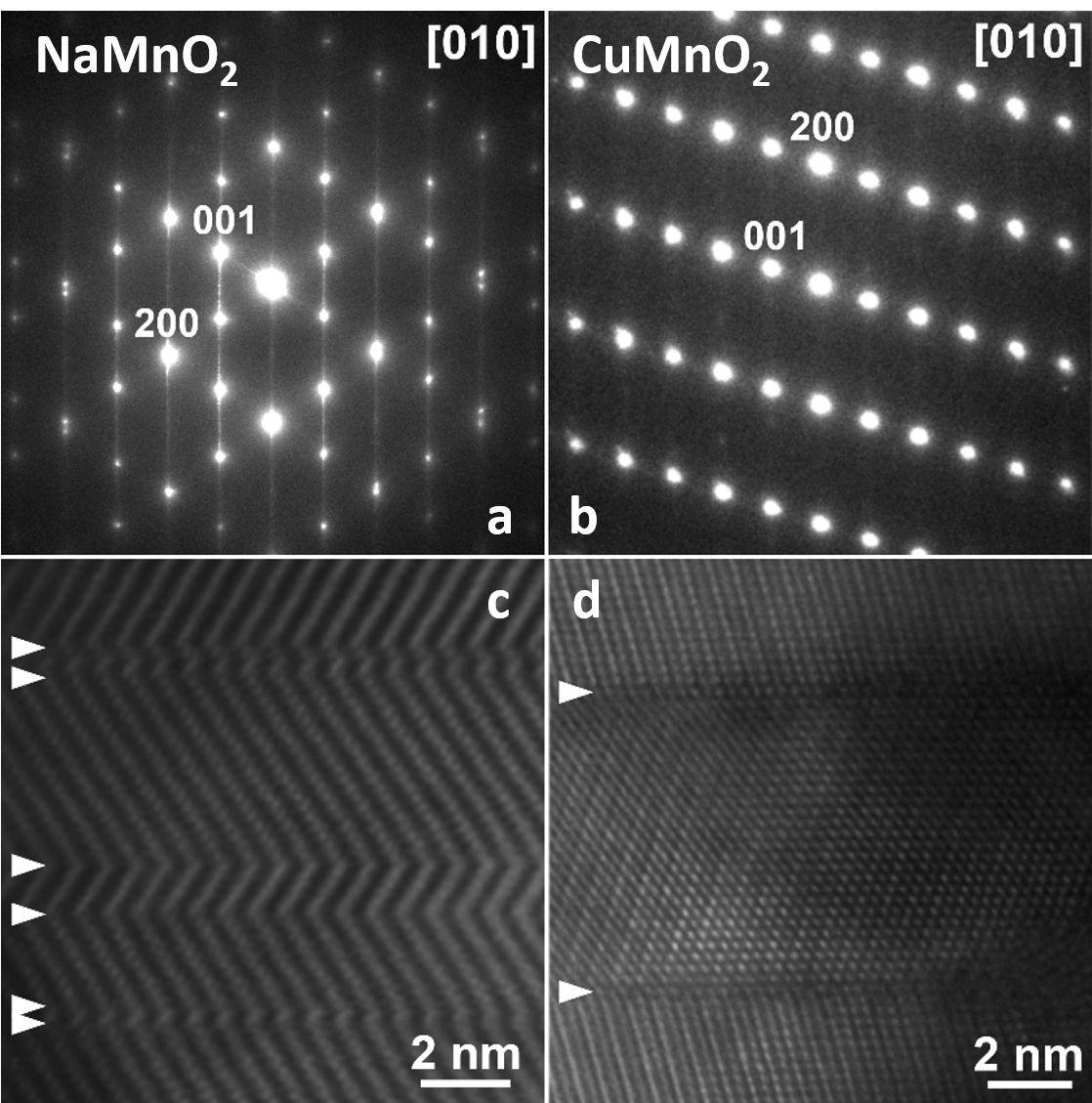}
	\caption{\label{fig:HRTEM} High-resolution transmission electron micrographs of \NMO\ and \CMO. (a) [010] electron diffraction pattern for \NMO, showing significant diffuse scattering due to defects in the structure. (b) [010] electron diffraction pattern for \CMO, with significantly less diffuse scattering, indicating fewer defects. (c) High-angle annular dark field scanning transmission electron microscopy (HAADF-STEM) image of \NMO. The white arrowheads indicate planar defects. (d) Same as (c), but for \CMO. Fewer defects are present.}
	
\end{figure}

High-resolution transmission electron microscopy (HRTEM) studies were performed at the Electron Microscopy for Materials Science (EMAT) facilities at the University of Antwerp as part of the ESTEEM2 infrastructure consortium. The \NMO\ sample was stored in an Ar-filled glove box. The TEM specimen was prepared in the glove box by pulverization of the polycrystalline specimen in a mortar with anhydrous ethanol or n-hexane and depositing drops of the resulting suspension on holey carbon grids. The specimen was transported and inserted into the microscope under dry Ar, completely excluding contact with air. The \CMO\ TEM sample was prepared in a similar fashion, but since it is not air sensitive, the preparation was carried at ambient conditions. Electron diffraction (ED) patterns were obtained with a Tecnai G$^2$ electron microscope operating at 200~kV, while high-angle annular dark-field scanning transmission electron microscopy (HAADF-STEM) images were obtained with a Titan G$^3$ electron microscope operating at 300 kV. Fig.~S\ref{fig:HRTEM}(a) displays an electron diffraction pattern for \NMO. Strong diffuse scattering is observed between the Bragg peaks, indicating significant disorder in \NMO. A real-space HAADF-STEM image of \NMO\ is shown in Fig.~S\ref{fig:HRTEM}(c), revealing planar defects (marked by white arrowheads) forming twin planes and anti-phase boundaries as the source of the lines of diffuse scattering in the diffraction pattern. For example, 
the \textit{a} and \textit{c} lattice parameters are interchanged across the twin planes, disrupting the stacking coherence of the layers of MnO$_6$ octahedra~\cite{abaku;cm14}.
The corresponding electron diffraction pattern and real-space HAADF-STEM image for \CMO\ are shown in Fig.~S\ref{fig:HRTEM}(b,d). The lines of diffuse scattering are significantly weaker than those in \NMO, and the real-space image reveals a lower density of planar defects. Thus, the HRTEM characterization indicates that \NMO\ has a greater degree of defect-induced structural disorder than \CMO. 

%